\documentclass[12pt]{article}
\usepackage{amsmath}
\usepackage{cite}
\usepackage[dvipdfmx]{graphicx}
\begin{document}    
\begin{flushright}
KANAZAWA-19-03\\
June, 2019
\end{flushright}
\vspace*{1cm}

\renewcommand\thefootnote{\fnsymbol{footnote}}
\begin{center} 
  {\Large\bf Pati-Salam unification with a spontaneous $CP$ violation}
\vspace*{1cm}

{\Large Daijiro Suematsu}\footnote[1]{e-mail:
~suematsu@hep.s.kanazawa-u.ac.jp}
\vspace*{0.5cm}\\

{\it Institute for Theoretical Physics, Kanazawa University, 
Kanazawa 920-1192, Japan}
\end{center}
\vspace*{1.5cm} 

\noindent
{\Large\bf Abstract}\\
Recent neutrino oscillation experiments suggest that the PMNS matrix
in the lepton sector has a $CP$ violating phase as the CKM matrix
in the quark sector. However, origin of these phases in both matrices are 
not clarified by now. Although complex Yukawa couplings could induce 
these phases, they remain as free parameters of the model even in that case. 
If the $CP$ symmetry is considered to be spontaneously broken, they are 
expected to be determined by some physics at a much lower energy scale 
than the Planck scale. We study such a possibility in a framework of 
Pati-Salam type unification. We also discuss other phenomenological issues in it.

\newpage
\setcounter{footnote}{0}
\renewcommand\thefootnote{\alph{footnote}}

\section{Introduction}
A $CP$ violation in a quark sector has been confirmed to be explained
by a CKM phase through experiments of the $B$ meson system.
However, its origin is not known still now.
Although the CKM phase can be derived from complex Yukawa couplings
of quarks \cite{ckm},  the $CP$ symmetry is considered 
to be explicitly broken in such a case and then the CKM phase remains 
as a free parameter of the model.
Even if its origin could be explained in some physics at the Planck scale,
it seems to be difficult to confirm it through experiments.
As another problem related to the $CP$ violation, we have
strong $CP$ problem \cite{stcp}.
The experimental bound of the electric dipole moment of a neutron
suggests that $\bar\theta~{^<_\sim}~10^{-10}$ should be satisfied \cite{nedm},
where $\bar\theta$ is defined as
$\bar \theta\equiv \theta_{\rm QCD} +{\rm arg}(\det {\cal M}_u{\cal M}_d)$
for up and down type quark mass matrices ${\cal M}_u$ and ${\cal M}_d$.
Since a QCD parameter $\theta_{\rm QCD}$ and the second term caused
from the quark masses are irrelevant each other, the required smallness 
of $\bar\theta$ seems to be unnatural, which is called the strong 
$CP$ problem in the standard model (SM).

One of the solutions for this problem is known to be presented
by the Peccei-Qiunn (PQ) mechanism \cite{pq}. Since its validity
could be examined through the existence of a light pseudo scalar 
called axion \cite{axion,ksvz,dfsz}, axion search is now performed 
in various experiments \cite{asearch}.
As another solution for the strong $CP$ problem, 
the Nelson-Barr (NB) model is known \cite{nb}. 
In this scenario, the $CP$ symmetry is assumed to be an exact symmetry
and then $\theta_{\rm QCD}=0$ is satisfied.  
If quark mass matrices take a special form based on some symmetry
to satisfy ${\rm arg}(\det {\cal M}_u{\cal M}_d)=0$,
$\bar\theta=0$ could be realized at least at a tree-level 
even after the spontaneous $CP$ violation.
On the other hand, this spontaneous $CP$ violation could give
a $CP$ phase in the CKM matrix. In this point, 
the scenario is interesting since it could explain an origin of the $CP$
violation at a much lower energy scale than the Planck scale.
Moreover, if a $CP$ breaking sector couples also with leptons,
a $CP$ phase in the PMNS matrix \cite{pmns}, whose existence is suggested
through the long baseline neutrino oscillation experiments such as
NOvA and T2K, might be explained by the same source as the quark sector.

A minimal simple example of the NB type model has been proposed 
by Bento, Branco and Parada (BBP) \cite{bbp}.
In this model, extra heavy vector-like down type quarks are introduced,
and a $Z_2$ symmetry is imposed on the model which controls 
the down type quark mass matrix so as not to bring about 
a contribution to $\bar\theta$ through ${\rm arg}(\det {\cal M}_u{\cal M}_d)$
after the spontaneous $CP$ violation. 
If we impose a global $U(1)$ symmetry instead of the $Z_2$ symmetry 
and assign its charge to these extra heavy quarks,
it is easy to find that the required form of the mass matrix could be 
realized in the same way.    
In such a case, interestingly enough, the model has the similar structure to
an invisible axion model by KSVZ \cite{ksvz}, which solves the
strong $CP$ problem through the PQ mechanism.
If the introduced global $U(1)$ works as the PQ 
symmetry, the contribution to $\bar\theta$ through radiative corrections 
to ${\rm arg}(\det {\cal M}_u{\cal M}_d)$ could be erased out. 
In that case, one of the problems in the NB model which are 
pointed out in \cite{dd} could disappear.
In this paper, we study this scenario in a Pati-Salam type
unified model, in which the $CP$ phases in the CKM matrix and
the PMNS matrix could be related.    

The remaining part of the paper is organized as follows.
In section 2, we introduce our model and discuss a possible origin of
$CP$ phases in both the CKM and PMNS matrices.
The generation of small neutrino masses is also addressed.
We additionally examine a possible spontaneous $CP$ violation in the model.
In section 3, we discuss several phenomenological issues in the model.
Section 4 is devoted to the summary of the paper. 

\section{Origin of $CP$ violation}
\subsection{A Pati-Salam type unified model}
We consider a unification model of quarks and leptons
via Pati and Salam \cite{ps}. The gauge symmetry is taken to be 
$SU(4) \times SU(2)\times U(1)_X$ in which the forth color is
identified with a lepton.
Fermion contents and their representations under this gauge group
are assumed to be
\begin{equation}
f_{L_i}~(4, 2, 0), \quad h_{R_i}~(4, 1, 1/2), \quad k_{R_i}~(4, 1, -1/2), 
\end{equation}
where $i$ is the generation index $(i=1, 2, 3)$.
As easily found, these contain all ordinary quarks and leptons.
We also introduce additional vector-like colored fermions
$F_{L,R}~(4, 1, -1/2)$, and $n$ triplet fermions
$\Sigma_{R_\alpha}~(1, 3, 0)$ where $\alpha=1 - n$ and they are
defined as
\begin{equation}
\Sigma_{R_\alpha}\equiv\sum_{a=1}^3\frac{\tau^a}{2}\Sigma^a_{R_\alpha}
=\frac{1}{2}\left(\begin{array}{cc}
\Sigma^0_{R_\alpha}&\sqrt 2\Sigma^+_{R_\alpha}\\
\sqrt 2\Sigma^-_{R_\alpha} & -\Sigma^0_{R_\alpha} \\
\end{array}\right).
\end{equation}
On the other hand, scalar contents and their representations are taken
to be
\begin{eqnarray}
 && \Phi~(4, 1, 1/2), \quad \Psi~(4, 1, 1/2), \quad \phi~(1, 2, -1/2), \quad 
\eta ~(1, 2, -1/2),  \nonumber \\
&& \sigma~(1, 1, 0),  \quad S~(1, 1, 0), \quad s~(1,1,0).
\end{eqnarray}
In addition to this structure, we impose a global $U(1)\times Z_8$ symmetry.
Its charge is assigned to these
fields as follows,
\begin{eqnarray}
&&  f_{L_i}, ~h_{R_i}, ~k_{R_i} \Rightarrow (0, 1), \quad   F_L, \Rightarrow (0, 7), \quad 
F_R \Rightarrow (2, 1), \quad \Sigma_{R_\alpha}\Rightarrow (1, 1), \quad 
S \Rightarrow (0, 6), \nonumber \\ 
&& \sigma \Rightarrow (2, 2), \quad  \eta\Rightarrow (-1, 1),  
\quad \Phi \Rightarrow (0, 4), \quad \Psi, ~\phi \Rightarrow (0, 0), \quad 
s \Rightarrow (0, 1).   
\label{gl2}
\end{eqnarray}
We also assume that $CP$ is an exact symmetry of the model.
Although $\Sigma_{R_\alpha}$ and $\eta$ might be considered needless in the model 
for the explanation of features shown through several experiments 
which cannot be explained 
in the SM framework,\footnote{If the axion is identified with the dark matter, 
they might be needless. However, we would like to consider much wider 
possibilities because of reasons which are addressed later.} 
we start our discussion in these field contents. 

If we adopt these field contents,
Yukawa couplings invariant under the imposed symmetry are written as
\begin{equation}
  -{\cal L}_y= y_{ij}^h\bar f_{L_i}\phi h_{R_j}
  +y_{ij}^k\bar f_{L_i}\tilde\phi k_{R_j}
  +y_iS\bar F_L k_{R_i}   + x\sigma^\ast \bar F_LF_R
+\gamma_{\Sigma_\alpha}\sigma^\ast\bar\Sigma_{R_\alpha}^c\Sigma_{R_\alpha}
+  {\rm h.c.}, 
\label{yukawa} 
\end{equation}
where $\tilde\phi=i\tau_2\phi^\ast$.
On the other hand, scalar potential is expressed as
\begin{eqnarray}
V&=&\tilde m_S^2(S^\dagger S)+ \tilde m_\sigma^2(\sigma^\dagger \sigma)+
\tilde m_s^2(s^\dagger s)+
\kappa_S(S^\dagger S)^2 +\kappa_\sigma(\sigma^\dagger \sigma)^2+
\kappa_s(s^\dagger s)^2 
+\kappa_{S\sigma}(S^\dagger S)(\sigma^\dagger\sigma)  \nonumber\\
&+&\kappa_{s\sigma}(s^\dagger s)(\sigma^\dagger\sigma)+
\kappa_{Ss}(S^\dagger S)(s^\dagger s)
+\kappa_{\sigma\phi}(\sigma^\dagger\sigma)(\phi^\dagger\phi)
+\kappa_{S\phi}(S^\dagger S)(\phi^\dagger\phi) 
+\kappa_{s\phi}(s^\dagger s)(\phi^\dagger\phi)  \nonumber\\
&+&\kappa_{\sigma\eta}(\sigma^\dagger\sigma)(\eta^\dagger\eta)
+\kappa_{S\eta}(S^\dagger S)(\eta^\dagger\eta) 
+\kappa_{s\eta}(s^\dagger s)(\eta^\dagger\eta)  \nonumber\\
&+&\tilde m_\phi^2(\phi^\dagger\phi) +\tilde m_\eta^2(\eta^\dagger\eta)+ 
\lambda_1(\phi^\dagger\phi)^2+\lambda_2(\eta^\dagger\eta)^2
+\lambda_3(\phi^\dagger\phi)(\eta^\dagger\eta)  
+\lambda_4(\phi^\dagger\eta)(\eta^\dagger\phi)  \nonumber \\
&+&m_\Phi^2(\Phi^\dagger\Phi)+m_\Psi^2(\Psi^\dagger\Psi)+
\zeta_1(\Phi^\dagger\Phi)^2+ \zeta_2(\Psi^\dagger\Psi)^2+
\zeta_3(\Phi^\dagger\Phi)(\Psi^\dagger\Psi)+
\zeta_4(\Phi^\dagger\Psi)(\Psi^\dagger\Phi)    \nonumber\\
&+&(\zeta_\sigma\sigma^\dagger\sigma+\zeta_SS^\dagger S 
+\zeta_ss^\dagger s +\zeta_\phi\phi^\dagger\phi+\zeta_\eta\eta^\dagger\eta)
(\Phi^\dagger\Phi +\Psi^\dagger\Psi)  \nonumber   \\
&+& V_b(S, S^\dagger, \sigma^\dagger\sigma, s^\dagger s, 
\Phi^\dagger\Psi, \Psi^\dagger\Phi, \phi^\dagger\phi,
\eta^\dagger\eta),
\label{pot}
\end{eqnarray}
where $V_b$ contains potential terms which are invariant under
the symmetry mentioned above but it violates the $S$ number conservation.
Since $CP$ is assumed to be exact, all coupling constants are real.
If $\Phi$ and $\Psi$ get vacuum expectation values (VEVs) such as
$\langle\Phi\rangle=\langle\Psi\rangle=(0, 0, 0, \Lambda)^T$ 
for example,\footnote{$\langle\Phi\rangle=\langle\Psi\rangle$ is assumed
just for simplicity.}
the gauge symmetry is broken to the one of the SM
\begin{equation}
  SU(4)\times SU(2)\times U(1)_X \quad\stackrel{\langle\Phi\rangle, 
\langle\Psi\rangle}
  \longrightarrow \quad SU(3)_C\times SU(2)_L\times U(1)_Y.
\end{equation}
The weak hypercharge $U(1)_Y$ whose charge is normalized as 
$Q_{EM}=\frac{\tau_3}{2}+Y$ is obtained as a linear combination of
a diagonal generator $T_{15}$ of $SU(4)$ and a charge $X$ of $U(1)_X$ as
\begin{equation}
Y=\frac{2}{\sqrt 6}T_{15} + X,
\end{equation}
where $T_{15}=\frac{1}{2\sqrt{6}}{\rm diag}(1,1,1,-3)$.
We note that the imposed global $U(1)$ symmetry remains unbroken
but $Z_8$ is broken to $Z_4$ at this stage. 
All fermions remain massless since they have no Yukawa couplings only 
with $\Phi$ and $\Psi$.

After this symmetry breaking, each fermion is decomposed to the
contents of the SM such as
\begin{equation}
f_{L_i}=(q_{L_i}, \ell_{L_i}), \quad h_{R_i}=(u_{R_i}, N_{R_i}), 
\quad k_{R_i}=(d_{R_i}, e_{R_i}), 
\end{equation}
where $q_{L_i}$ and $\ell_{L_i}$ are $SU(2)_L$ doublet quarks and leptons, and 
$u_{R_i}$, $d_{R_i}$ and $e_{R_i}$ are singlet quarks and 
charged leptons, respectively.
The vector-like fermions $F_{L,R}$ are decomposed as $(D_{L,R}, E_{L,R})$.
If we use these decomposed fermions, 
Yukawa couplings in eq.~(\ref{yukawa}) are expressed 
as\footnote{We note that a non-renormalizable operator such as
$S^\ast(\Psi\bar F_L)(\Phi^\dagger k_{R_i})$ which is invariant under 
the imposed symmetry induces the Yukawa terms $S^\ast\bar D_Ld_{R_i}$ and 
$S^\ast\bar E_Le_{R_i}$.}   
\begin{eqnarray}
  -{\cal L}_y&=& y_{ij}^u\bar q_{L_i}\phi u_{R_j}
  + y_{ij}^d\bar q_{L_i}\tilde\phi d_{R_j}
+(y_i^DS+\tilde y_i^{D} S^\ast)\bar D_L d_{R_i}   + x_D\sigma^\ast \bar D_LD_R 
\nonumber \\
&+&y_{ij}^\nu\bar \ell_{L_i}\phi N_{R_j}+ y_{ij}^e\bar \ell_{L_i}\tilde\phi e_{R_j} 
+(y_i^ES+\tilde y_i^{E} S^\ast)\bar E_L e_{R_i}   + x_E\sigma^\ast \bar E_LE_R
\nonumber \\ 
&+&\gamma_{\Sigma_\alpha}\sigma^\ast\bar\Sigma_{R_\alpha}^c\Sigma_{R_\alpha}
+{\rm h.c.},
\label{lyukawa}
\end{eqnarray}
where the Yukawa coupling constants are expected to 
satisfy the conditions 
\begin{eqnarray}
y_{ij}^h=y_{ij}^u=y_{ij}^\nu, \quad y_{ij}^k=y_{ij}^d=y_{ij}^e, \quad
y_i=y_i^D=y_i^E, \quad \tilde y_i=\tilde y_i^D=\tilde y_i^E, \quad
x=x_D=x_E,
\label{inityuk}
\end{eqnarray}
at a unification scale $\Lambda$.
After the spontaneous breaking of $SU(4)$ via 
$\langle\Phi\rangle$ and $\langle\Psi\rangle$, new Yukawa couplings are expected 
to be induced effectively as invariant ones under 
the remaining symmetry,\footnote{It should be noted that the Yukawa term
$S^\ast\bar N^c_{R_i}N_{R_i}$ can be induced by a non-renormalizable operator
$S^\ast(\Phi^\dagger h_{R_i})(\Psi h_{R_i})$ invariant under the 
imposed symmetry, for example.}
\begin{eqnarray}
-{\cal L}^\prime_y&=& \left(y^N_iS+\tilde y^N_iS^\ast
+a_i\frac{s^2}{\Lambda}+\tilde a_i\frac{s^{\ast 2}}{\Lambda}\right)\bar N_{R_i}^cN_{R_i}
+\tilde h_{i\alpha}\frac{s^\ast}{\Lambda}\bar\ell_{L_i}\Sigma_{R_\alpha}\eta
\nonumber \\
&+& \left(b_i\frac{s^2}{\Lambda}+b_i\frac{s^{\ast 2}}{\Lambda}\right)\bar D_Ld_{R_i}
+\left(c_i\frac{s^2}{\Lambda}+c_i\frac{s^{\ast 2}}{\Lambda}\right)\bar E_Le_{R_i}
 +  {\rm h.c.},
\label{lyukawa2}
\end{eqnarray}
where we list up the terms up to dimension five.
The couplings $y^N_i$ and $\tilde y^N_i$ are assumed to be diagonal.
We also note that there is a nonrenormalizable dimension five operator
$\tilde\lambda_5\frac{\sigma}{\Lambda}(\phi^\dagger\eta)^2$ 
as an invariant one.
It plays a crucial role in the small neutrino mass generation as seen later.

In this effective model, we consider symmetry breaking due to
VEVs of the singlet scalars $\sigma$, $S$ and $s$ 
such as\footnote{The global symmetry $U(1)$ is broken to 
$Z_2$ by these VEVs. The $Z_2$ guarantees the stability of DM
as discussed later. }
\begin{equation}
\langle\sigma\rangle=we^{i\chi}, \qquad \langle S\rangle=ue^{i\rho}, \qquad
\langle s\rangle=ve^{i\psi}.
\label{svev}
\end{equation}
They could also break the $CP$ symmetry spontaneously.
Although we will discuss whether this spontaneous $CP$ violation
could be realistic or not in the present model later, we assume it for a while.
Here we note that
for $\bar D_Ld_{R_i}$, $\bar E_Le_{R_i}$ and $\bar N^cN_{R_i}$ 
in eqs.~(\ref{lyukawa}) and (\ref{lyukawa2}) there are contributions 
from the dimension four and five operators. 
We can expect that the formers give the dominant contribution 
as long as $v~{^<_\sim}~u$ is satisfied at least.
We suppose such a situation and take account of these contributions 
only in the following study.
 
After this symmetry breaking, the potential for the remaining scalars
$\phi$ and $\eta$ can be written as
\begin{eqnarray}
V&=&m_\phi^2(\phi^\dagger\phi) +m_\eta^2(\eta^\dagger\eta)+ 
\lambda_1(\phi_1^\dagger\phi)^2+\lambda_2(\eta^\dagger\eta)^2
+\lambda_3(\phi^\dagger\phi)(\eta^\dagger\eta) \nonumber \\
&+&\lambda_4(\phi^\dagger\eta)(\eta^\dagger\phi)+
\frac{\lambda_5}{2}\left[(\phi^\dagger\eta)^2 +{\rm h.c.}\right],
\label{spot}
\end{eqnarray}
where $\lambda_5$ is defined as $\lambda_5=\tilde\lambda_5\frac{w}{\Lambda}$ 
and it is real.\footnote{The $CP$ phase $\chi$ can be removed by the field redefinition of $\eta$. It changes 
$h_{i\alpha}$ in eq.~(\ref{lyukawa2}) to $h_{i\alpha}e^{-i\frac{\chi}{2}}$.} 
The scalar masses are shifted through the symmetry
breaking effect as
\begin{equation}
  m_\phi^2= \tilde m_\phi^2+\kappa_{\sigma\phi}w^2
  +\kappa_{S\phi}u^2+\kappa_{s\phi}v^2+2\zeta_\phi\Lambda^2, \quad
  m_\eta^2= \tilde m_\eta^2+\kappa_{\sigma\eta}w^2
  +\kappa_{S\eta}u^2+\kappa_{s\eta}v^2+2\zeta_\eta\Lambda^2.  
\end{equation}  
Since $m_\phi$ and $m_\eta$ are supposed to take much smaller values 
than $\Lambda$,
serious fine tunings are required. However, we do not treat this hierarchy
problem in the present study and just assume that both
$m_\phi$ and $m_\eta$ are of $O(1)$ TeV.
The coupling constants $\lambda_i$ are also related to the ones at high
energy regions through threshold corrections at each symmetry
breaking scales \cite{thr}. 

An interesting feature of the present model is that
the spontaneous $CP$ violation through eq.~(\ref{svev}) could
derive both $CP$ phases in the CKM matrix and the PMNS matrix
keeping $\bar\theta=0$.
In the next part, we discuss how the $CP$ phases in both CKM and PMNS 
matrices are induced.

\subsection{A $CP$ phase in the CKM matrix}
The $CP$ symmetry is assumed to be exact in the model and 
then all the coupling constants in the Lagrangian are real.  
Thus, we cannot expect any origin of
$CP$ violation in the up type quark sector, which has no extended structure
compared with the SM. Since the up sector mass matrix
$m^u_{ij}=y^u_{ij}\langle\phi\rangle$ is real, they can be diagonalized by
orthogonal transformations $u_L^\prime=O^Lu_L$ and $u_R^\prime=O^Ru_R$.
In the present effective model,
on the other hand, we find that the down type quark sector 
has the same structure as the BBP model \cite{bbp}.
The BBP model is an extension of the SM by extra colored vector-like down 
type heavy quarks $(D_L, D_R)$ and a singlet complex scalar $S$.
We can apply their discussion to the present model to show how 
the $CP$ phase could be induced in the CKM matrix. 
Although the $Z_2$ symmetry is imposed to control the mass matrix 
in their model, the global $U(1)$ symmetry in eq.~(\ref{gl2}) could play 
the same role as it in the present model. 
Moreover, since this $U(1)$ is chiral and has a color anomaly, it can play
a role as the PQ symmetry which has a domain wall number one 
as in the KSVZ model \cite{ksvz}.
As a result, a Nambu-Goldstone boson produced as a result of its spontaneous 
breaking through the VEV $\langle\sigma\rangle$ could work as an axion
to solve the strong $CP$ problem without inducing the domain wall 
problem \cite{dw}.
On the other hand, since the axion phenomenology constrains a breaking
scale of this symmetry, we have to fix the scale $w$ to be \cite{fa}
\begin{equation}
10^9~{\rm GeV}< w <10^{12}~{\rm GeV}.
\label{fa}
\end{equation}  

The Yukawa couplings of the down type quarks shown in eq.~(\ref{lyukawa}) 
derive a $4\times 4$ mass matrix ${\cal M}_d$ as
\begin{equation}
(\bar d_{Li}, \bar D_L)\left(
\begin{array}{cc}
m^d_{ij} & 0 \\ {\cal F}^d_j & \mu_D \\
\end{array}\right)
\left(\begin{array}{c} d_{R_j} \\ D_R \\ \end{array} \right). 
\label{dmass}
\end{equation}
where $m_{ij}^d=y_{ij}^d\langle\tilde\phi\rangle$, 
${\cal F}_j^d=(y_j^D ue^{i\rho} + \tilde y_j^D 
ue^{-i\rho})$ and $\mu_D=x_Dwe^{i\chi}$.  
Due to the PQ mechanism, 
$\bar\theta=\theta_{QCD}+{\rm arg}(\det{\cal M}_u{\cal M}_d)=0$ 
is satisfied via the axion even after we take account of 
radiative corrections including the phases caused by 
the spontaneous $CP$ violation. 
Next we see that this phase can generate the CKM phase following
the BBP model.

We consider the diagonalization of a matrix ${\cal M}_d{\cal M}_d^\dagger$ 
by a unitary matrix such as
\begin{equation}
\left(\begin{array}{cc} A & B \\ C& D \\\end{array}\right)
\left(\begin{array}{cc} m^dm^{d\dagger} & m^d{\cal F}^{d\dagger} \\ 
  {\cal F}^dm^{d\dagger} & \mu_D\mu_D^\dagger +{\cal F}^d{\cal F}^{d\dagger} \\
\end{array}\right)
\left(\begin{array}{cc} A^\dagger & C^\dagger \\ B^\dagger 
& D^\dagger \\\end{array}\right)=
\left(\begin{array}{cc} m^2 & 0 \\ 0 & M^2 \\\end{array}\right),
\label{mass}
\end{equation}
where a $3\times 3$ matrix $m^2$ is diagonal in which the generation 
indices are abbreviated. Eq.~(\ref{mass}) requires
\begin{eqnarray}
   && m^dm^{d\dagger}=A^\dagger m^2A+ C^\dagger M^2C, \qquad
  {\cal F}^dm^{d\dagger}=B^\dagger m^2A+ D^\dagger M^2C, \nonumber \\
   && \mu_D\mu_D^\dagger+{\cal F}^d{\cal F}^{d\dagger}=
B^\dagger m^2 B+ D^\dagger M^2 D.
\end{eqnarray}  
If $\mu_D\mu_D^\dagger+{\cal F}^d{\cal F}^{d\dagger}$ is much larger than each 
components of ${\cal F}^dm^{d\dagger}$, which means
$u, w\gg \langle \tilde\phi\rangle$,
we find that $B, C$ and $D$ can be approximated as
\begin{equation}
  B\simeq -\frac{Am^d{\cal F}^{d\dagger}}{\mu_D\mu_D^\dagger
+{\cal F}^d{\cal F}^{d\dagger}},
 \qquad C\simeq\frac{{\cal F}^d m^{d\dagger}}{\mu_D\mu_D^\dagger
+{\cal F}^d{\cal F}^{d\dagger}},
   \qquad D\simeq 1,
\end{equation}
which guarantee the approximate unitarity of the matrix $A$. 
In such a case, it is also easy to find that
\begin{equation}
A^{-1}m^2A= m^dm^{d\dagger} -\frac{1}{\mu_D\mu_D^\dagger
+{\cal F}^d{\cal F}^{d\dagger}}
(m^d{\cal F}^{d\dagger})({\cal F}^dm^{d\dagger}).
\label{ckm}
\end{equation}
The right-hand side is an effective mass matrix of the ordinary 
down type quarks which is derived through the mixing with the extra heavy quarks. 
Since the second term can have complex phases in off-diagonal 
components as long as $y^D_i\not=\tilde y^D_i$ is satisfied, 
the matrix $A$ could be complex. 
Moreover, if $\mu_D\mu_D^\dagger<{\cal F}^d{\cal F}^{d\dagger}$ is satisfied,
the complex phase in $A$ could have a substantial magnitude
since the second term is comparable with the first term.
Since the CKM matrix is determined as $V_{CKM}={O^L}^TA$,
the $CP$ phase of $V_{CKM}$ is caused through the one of $A$.
Here, we have to note whether such phases could be physical or not 
is dependent on the flavor structure of Yukawa coupling $y^d$, $y^D$ 
and $\tilde y^D$. 
It should also be noted that the matrix $A$ needs to take an almost 
diagonal form as long as
there is no correlation between $A$ and $O^L$
since $V_{CKM}$ has a nearly diagonal form.
It may be instructive to show how the physical phase could be induced
through this mechanism using a concrete example.
We give such an example in Appendix.

\subsection{Neutrino masses and the PMNS matrix}
In the lepton sector, we can treat the charged lepton sector 
in the same way as the down type quark sector.
In fact, the Yukawa couplings in eq.~(\ref{lyukawa}) induce the charged   
lepton mass matrix as follows,
\begin{equation}
(\bar e_{Li}, \bar E_L)\left(
\begin{array}{cc}
m^e_{ij} & 0 \\ {\cal F}^e_j & \mu_E \\
\end{array}\right)
\left(\begin{array}{c} e_{R_j} \\ E_R \\ \end{array} \right),
\end{equation}
where $m_{ij}^e=y_{ij}^e\langle\tilde\phi\rangle$, 
${\cal F}_j^e=(y_j^E ue^{i\rho} + \tilde y_j^E ue^{-i\rho})$ and 
$\mu_E=x_Ewe^{i\chi}$. Since the mass matrix takes the same form as the one of
the down type quarks (\ref{mass}), the diagonalization matrix $\tilde A$ 
for the above charged lepton mass matrix could be complex and 
it should satisfy the relation
\begin{equation}
\tilde A^{-1}\tilde m^2\tilde A= m^em^{e\dagger} -\frac{1}{\mu_E\mu_E^\dagger
+{\cal F}^e{\cal F}^{e\dagger}}
(m^e{\cal F}^{e\dagger})({\cal F}^em^{e\dagger}),
\label{ckm}
\end{equation}
where $\tilde m^2$ corresponds to the diagonalized mass matrix $m^2$ in
eq.~(\ref{mass}).
As long as $\mu_E\mu_E^\dagger <{\cal F}^e{\cal F}^{e\dagger}$
is satisfied, non-negligible $CP$ phases could be expected in $\tilde A$
in the same way as the down type quark sector. 

On the other hand, small neutrino masses are expected to be 
produced not only by the type I seesaw \cite{seesaw} but also 
by the scotogenic type III seesaw \cite{scot3} in this model.
In fact, the lepton sector of the model has the structure
in which the scotogenic type III seesaw mechanism could work as found from 
the terms contained in eqs.~(\ref{lyukawa}) and (\ref{lyukawa2}).
Diagrams which contribute to the neutrino mass generation
are shown in Fig.1. 

\input epsf
\begin{figure}[t]
\begin{center}
\epsfxsize=14cm
\leavevmode
\epsfbox{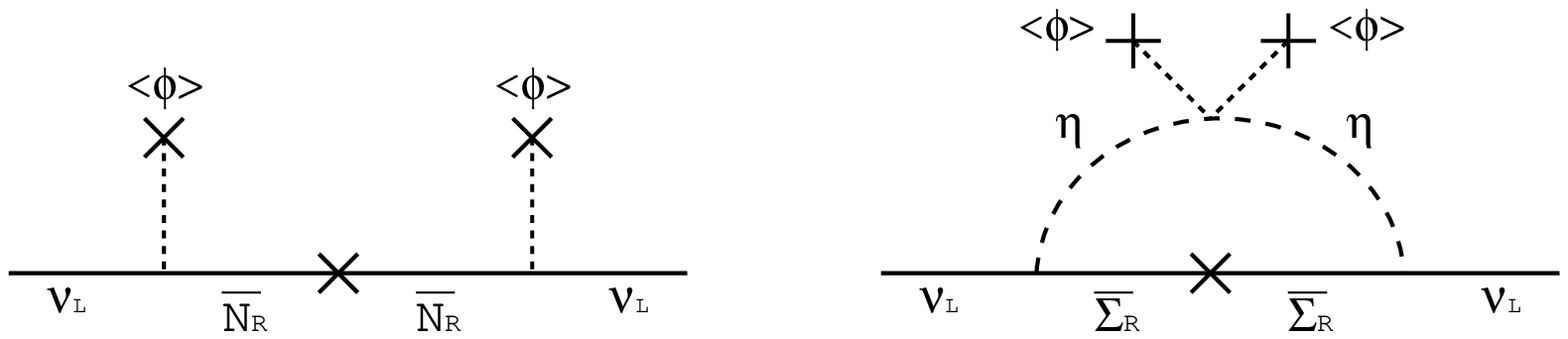}
\end{center}
\vspace*{-13mm}
    {\footnotesize {\bf Fig.~1} Left : A diagram for the neutrino mass generation
      due to the type I seesaw
      in the minimal model. Right : A one-loop diagram for the neutrino mass
      generation due to the scotogenic type III seesaw in the extended model.}  
\end{figure}  

\noindent
(a) Neutrino masses due to the type I seesaw\\
The singlet fermions $N_{R_i}$ get Majorana mass via the VEV
$\langle S\rangle$. On the other hand, they have Yukawa couplings with 
the doublet leptons and the ordinary Higgs doublet scalar
$\phi$. Thus, the ordinary type I seesaw makes neutrinos $\nu_{L_i}$
massive through the diagram shown in the left of Fig.~1. 
Neutrino mass matrix caused by this can be written as 
\begin{equation}
(\bar\nu_{L}^c, \bar N_{R}^c)
\left(\begin{array}{cc}
0& y^\nu \langle \phi\rangle \\    
y^{\nu T} \langle\phi\rangle  & y^N ue^{i\rho} + \tilde y^N ue^{-i\rho} \\ \end{array}\right)
\left(\begin{array}{c} \nu_L  \\ N_R \\ \end{array}\right).
\end{equation}
Since $u\gg \langle\phi\rangle$ is supposed in this model,
the contribution to neutrino masses from this diagram is estimated as
\begin{equation}
{\cal M}_{ij}^{(a)}=\sum_{k=1}^3y^\nu_{ik}y_{jk}^{\nu}
\frac{\langle\phi\rangle^2}{y^N_kue^{i\rho}+\tilde y^N_kue^{-i\rho}}.
\end{equation} 
The neutrino Yukawa couplings $y_{ik}^\nu$ satisfy the same relation 
as the Yukawa couplings of the up type quarks as found in eq.~(\ref{inityuk}).
Since $\tilde A$ is expected to take an almost diagonal form as $A$,
the PMNS matrix is consider to have the similar form as the CKM matrix.
This means that other contributions to the neutrino masses are indispensable
for the explanation of large flavor mixing required by the neutrino 
oscillation data.
This is one of the reasons why we consider the extended structure with 
$\eta$ and $\Sigma_{R_\alpha}$.
These fields could give additional contributions to the neutrino masses
in the following way.

\noindent
(b) Neutrino masses due to the scotogenic type III seesaw\\
As found in eq.~(\ref{lyukawa2}), $\Sigma_{R_\alpha}$ has Yukawa couplings
with $\nu_{L_i}$. However, since $\phi$ has no coupling with these and
$\eta$ is assumed to have no VEV,
neutrino masses via $\Sigma_{R_\alpha}$ are not generated
at a tree level but generated at a one-loop level.
The coupling $\frac{\lambda_5}{2}(\eta^\dagger\phi)^2+{\rm h.c.}$ 
brings about a small
mass difference between the real and imaginary components of $\eta^0$. 
As its result, the one-loop diagram shown in the right of Fig.~1 gives
a contribution to the neutrino masses.   
It can be estimated as
\begin{eqnarray}
  {\cal M}_{ij}^{(b)}&=&\sum_{\alpha=1}^{n_\Sigma}
  \frac{h_{i\alpha}h_{j\alpha}\lambda_5\langle\phi\rangle^2
e^{-i\rho}}{32 \pi^2M_{\Sigma_\alpha}}
  \left[\frac{M_{\Sigma_\alpha}^2}{M_\eta^2-M_{\Sigma_\alpha}^2}
    \left(1+\frac{M_{\Sigma_\alpha}^2}{M_\eta^2-M_{\Sigma_\alpha}^2}
    \ln\frac{M_{\Sigma_\alpha}^2}{M_\eta^2}\right) \right]  \nonumber \\
  &\simeq&
  \sum_{\alpha=1}^{n_\Sigma}
  \frac{h_{i\alpha}h_{j\alpha}\lambda_5\langle\phi\rangle^2
e^{-i\rho}}{32 \pi^2M_{\Sigma_\alpha}}
  \ln\frac{M_{\Sigma_\alpha}^2}{M_\eta^2}.
\end{eqnarray}
where $M_{\Sigma_\alpha}=\gamma_{\Sigma_\alpha}w$ and
$M_\eta^2=m_\eta^2+(\lambda_3+\lambda_4)\langle\phi\rangle^2$.
The second similarity is satisfied for $M_\eta=O(1)$ TeV since
$w$ is much larger than a TeV scale as discussed in the previous part.
Although neutrino mass eigenvalues are determined through
${\cal M}_{ij}^{\nu}={\cal M}_{ij}^{(a)}+{\cal M}_{ij}^{(b)}$,
${\cal M}_{ij}^{(a)}$ should be sufficiently small compared with ${\cal M}_{ij}^{(b)}$
for large flavor mixings. 
If we consider that this matrix is diagonalized by a unitary matrix
$U$ as $U^T{\cal M}^\nu U={\cal M}^{\rm diag}$, the PMNS matrix is obtained
as $V_{PMNS}=\tilde{A}^\dagger U$ which could have a Dirac phase and two
Majorana phases. An example of $V_{PMNS}$ obtained through this 
framework in a simple model is given in Appendix. 

Next, we address the constraint on the relevant parameters
caused by the neutrino oscillation data.
Since ${\cal M}^{(a)}$ should be a subdominant contribution to
the neutrino masses,
we have to extend the model at least with two triplet fermions 
$(n_\Sigma=2)$ for the explanation of the neutrino oscillation data. 
In order to estimate the required magnitude of the
neutrino Yukawa couplings in such a case, 
we suppose, for simplicity and definiteness, the tri-bimaximal flavor structure
for $h_{i\alpha}$ as \cite{tribi}
\begin{equation}
h_{e1}=0, ~ h_{\mu 1}=h_{\tau 1}\equiv h_1; \qquad 
h_{e2}=h_{\mu 2}=-h_{\tau 2}\equiv h_2,
\label{flavor}
\end{equation}
and also diagonal $y_{ij}^\nu$ such as $y_{ij}^\nu=y_i^\nu\delta_{ij}$ with
$y_1^\nu\ll y_2^\nu \ll y_3^\nu$.\footnote{This assumption is adopted 
due to the relation (\ref{inityuk}) to the up type quarks which is caused by the
$SU(4)$ symmetry.} 
We also assume $y^N_{1,2}=0$ and $\tilde y^N_3=0$, for simplicity.
Under this assumption, if the normal hierarchy for the neutrino masses
is assumed, squared mass differences required by the neutrino 
oscillation data suggest \cite{pdg}
\begin{eqnarray}
&&h_1^2\simeq 9.3\times 10^{-3}\left(\frac{10^{-2}}{\lambda_5}\right)
\left(\frac{M_{\Sigma_1}}{10^{10}~{\rm GeV}}\right)
\left[1-\left(\frac{y^\nu_3}{4.4\times 10^{-3}}\right)^2
\left(\frac{10^{10}~{\rm GeV}}{M_{N_3}}\right)\right], \nonumber \\
&&h_2^2\simeq 9.3\times 10^{-4}\left(\frac{10^{-2}}{\lambda_5}\right)
\left(\frac{M_{\Sigma_2}}{10^{10}~{\rm GeV}}\right), \qquad
y^{\nu 2}_1< 9.2\times 10^{-7} \left(\frac{M_{N_1}}{10^{9}~{\rm GeV}}\right),
\nonumber \\
&& y^{\nu 2}_2< 9.2\times 10^{-7} \left(\frac{M_{N_2}}{10^{9}~{\rm GeV}}\right),
\qquad
y^{\nu 2}_3< 1.0\times 10^{-5} \left(\frac{M_{N_3}}{10^{10}~{\rm GeV}}\right),
\label{nmcond}
\end{eqnarray}  
where $M_{N_{1,2}}=\tilde y^N_{1,2}u$ and $M_{N_3}=y^N_3u$,
and $M_\eta=1$~TeV is also assumed. 

Finally, it may be useful to present a remark on the extension by the 
vector-like fermions.
Although these fermions are introduced to the down sector
in the above discussion, the CKM phase could be derived in the same way
even if we introduce them to the up sector.
However, the situation could be largely changed for the $CP$ phases
in the PMNS matrix and the small neutrino mass generation. 
The present choice seems to be crucial for the present scenario.
It could also play an important role when we consider an
embedding of the model into a fundamental model 
at the Planck scale region.\footnote{The model might be embedded into an
effective model derived by a suitable compactification of $E_8\times E_8^\prime$
superstring \cite{ms}.}

\subsection{Spontaneous $CP$ violation}
In the previous part, we just assume that eq.~({\ref{svev}) is
  realized as a potential minimum. Here, we discuss in what situation the
  spontaneous $CP$ violation could occur in a realistic way
  in the present model. 
  The condition required for the spontaneous $CP$ violation has been
  studied in detail in \cite{vcp}.
If we follow their results, the VEVs of $\sigma$ is found not to break 
the $CP$ symmetry spontaneously,
  and then $\chi=0$. The reason is that the spurions
  for it cannot be introduced since the imposed global $U(1)$ symmetry 
is assumed to be
  exact except for the color anomaly effect.
  On the other hand, we can introduce the spurions
  for $S$ which has no global $U(1)$ charge.
 In fact, if we introduce the terms such as $S^4$ and $S^2$ which break
  a $U(1)$ symmetry corresponding to the $S$ number, a 
nonzero $\rho$ could appear as a potential minimum.\footnote{We do not 
consider such terms for $s$ in the present study.}

Relevant potential is found from eq.~(\ref{pot}) to be
\begin{equation}
V_{CP}=\bar m^2_S(S^\dagger S)+\bar m^2_\sigma(\sigma^\dagger\sigma)
+\kappa_S(S^\dagger S)^2 +\kappa_\sigma(\sigma^\dagger\sigma)^2
+\kappa_{S\sigma}(S^\dagger S)(\sigma^\dagger\sigma)+V_b,
\label{cppot}
\end{equation}
where $\bar m_a^2=\tilde m_a^2+\zeta_a\Lambda^2$ $(a=S, \sigma)$ 
and $\tilde m_a^2>0$ and $\zeta_a<0$ are assumed
since we suppose that the potential minimum is fixed as a result of the
$SU(4)$ breaking.  
As examples, we consider two cases for $V_b$ in (\ref{cppot}) 
such as\footnote{We note that terms proportional to $S^2$ are induced 
through the $SU(4)$ breaking from an operator $\Phi^\dagger\Psi S^2$ 
wihich is invariant under the imposed symmetry.}
  \begin{eqnarray} 
       &{\rm (i)}&~~ V_b=\alpha(S^4 +S^{\dagger 4}) + \mu^2(S^2 +S^{\dagger 2} ), 
     \nonumber  \\
       &{\rm (ii)}&~~ V_b=\alpha(S^4 +S^{\dagger 4} )+
   \beta(S^2+S^{\dagger 2})(\sigma^\dagger\sigma). 
  \end{eqnarray}
Here, we confine our study to the situation where 
the VEVs $u$ and $w$ are determined by a part
of $V_{CP}$ except for $V_b$.
It could be realized for $\kappa_S\gg \alpha$ 
and $|\bar m_S^2|\gg|\mu^2|$
in the case (i) and also for $\kappa_S\gg \alpha$ and $|\beta|\ll 1$ in
the case (ii).  
The potential minimum could be found for sufficiently small
$|\kappa_{S\sigma}|$ in both cases 
\begin{equation}
  u^2=-\frac{\bar m_S^2}{2\kappa_S}, \qquad
  w^2=-\frac{\bar m_\sigma^2}{2\kappa_\sigma}, 
\label{vev}
\end{equation}
and also the $CP$ phase is determined as
\begin{equation}
  {\rm (i)}~~ \cos 2\rho=-\frac{\mu^2}{4\alpha u^2},
  \qquad
      {\rm (ii)}~~\cos 2\rho=-\frac{\beta w^2}{4\alpha u^2},
\label{condcp}      
\end{equation}  
in each case.
These examples show that the spontaneous $CP$ violation
could occur through the scalar $S$ as long as suitable values of 
the parameters are chosen.
In fact, for example, if $\mu^2=-4\alpha u^2$ is satisfied for $\alpha\ll 1$
and $|\mu^2|\ll u^2$, the maximum $CP$ phase
$\rho\simeq \frac{\pi}{2}$ could be realized in the case (i).
We should note that these conditions on $\alpha$ and $\mu^2$ is
consistent with the requirement for which $u$ and $w$ are determined
as eq.~(\ref{vev}).
In the case (ii), the maximum $CP$ phase is obtained for
$\beta w^2\simeq 4\alpha u^2$ which is consistent with the determination
of $u$ and $w$ as found from eq.~(\ref{vev}).  
As a result of this symmetry breaking, the mass of $S$ is
fixed as $m_{S_R}=\sqrt{4\kappa_S} u$.

On the other hand, in order for this breaking to cause large $CP$ phases
 in both the CKM and PMNS matrices, the conditions
$\mu_D\mu_D^\dagger<{\cal F}^d{\cal F}^{d\dagger}$ and 
$\mu_E\mu_E^\dagger<{\cal F}^e{\cal F}^{e\dagger}$
should be satisfied as discussed before. They are supposed to require 
\begin{equation}
u~>~ w,
\label{vevc}
\end{equation}
as long as the relevant Yukawa couplings has a similar magnitude. 
This condition can be easily satisfied for suitable parameters
as found from eq.~(\ref{vev}).
Although the tuning of parameters is necessary, the present scenario
is found to work as long as the scalar potential takes a suitable form.
We can expect that the required $CP$ violation is induced in both the
quark and lepton sectors based on the same origin.

\section{Phenomenology}
In the previous part, we addressed that the $CP$ problem in the SM
could be solved in this model. In this section, we order several discussions
and comments on other phenomenological issues.   

\subsection{Inflation}
  The model has candidates for the inflaton such as $\sigma$ and $S$.
  They can have non-minimal couplings with Ricci scalar $R$ \cite{nonm-inf}
\begin{equation} 
  \frac{1}{2}\xi_\sigma\sigma^\dagger\sigma R, \qquad
  \frac{1}{2}\left[\xi_{S_1}S^\dagger S +
\frac{\xi_{S_2}}{2}(S^2+S^{\dagger 2})\right]R.
\end{equation}
Although the real and imaginary components of $\sigma$
have the same coupling $\xi_\sigma$, only the real part of $S$ could
have a nonzero coupling $\frac{1}{2}\xi S_R^2R$ 
in the case $\xi_{S_1}=\xi_{S_2}$ 
where $S\equiv\frac{1}{\sqrt 2}(S_R+iS_I)$ and $\xi\equiv \xi_{S_1}+\xi_{S_2}$.
If we suppose that a coupling $\xi$ takes a 
sufficiently large value in such a case, inflation via $S_R$ is expected to occur 
in the same way as the Higgs inflation \cite{higgsinf}.                 
A nice feature in this scenario is that the dangerous unitarity violation caused 
by a higher order mixing between $S_R$ and $S_I$ \cite{unitarity} 
is not  induced at $\frac{M_{\rm pl}}{\xi}$ but
could be suppressed 
at least up to an inflation scale $\frac{M_{\rm pl}}{\sqrt\xi}$ \cite{sinf}.

The potential of the inflaton can be expressed in the Einstein frame as
\begin{equation}
V_E=\frac{\kappa_S}
{\left(1+\frac{\xi S_R^2}{M_{\rm pl}^2}\right)^2}
\left[\frac{1}{2}(S_R^2+S_I^2)-u^2\right]^2.
\end{equation}
Since the canonically normalized inflaton $\chi$
is defined as
\begin{equation}
  \frac{d\chi}{d S_R}=\frac{1}{1+\frac{\xi S_R^2}{M_{\rm pl}^2}}
\left(1+\frac{\xi S_R^2}{M_{\rm pl}^2}
  +\frac{6\xi^2 S_R^2}{M_{\rm pl}^2}\right)^{1/2},
\end{equation}
$\chi$ and $S_R$ are related each other as
$S_R\propto\exp\frac{\chi}{\sqrt 6 M_{\rm pl}}$ at a large field region
$S_R^2\gg \frac{M_{\rm pl}^2}{\xi}$. 
In that region,  the potential of $\chi$
becomes constant $V_E=\frac{\kappa_S M_{\rm pl}^4}{4\xi^2}$ as long as
$S_R\gg S_I$ is satisfied.
The slow roll parameters for $\chi$ can be expressed as
\begin{equation}
  \epsilon\equiv\frac{M_{\rm pl}^2}{2}\left(\frac{V_E^\prime}{V_E}\right)^2
  =\frac{3}{4N_e^2}, \qquad
\eta\equiv M_{\rm pl}^2\frac{V_E^{\prime\prime}}{V_E}=-\frac{1}{N_e}
\end{equation}  
by using the e-foldings number $N_e$.
If we take $N_e=60$, we obtain the spectral index $n_s=0.97$ and 
the tensor-to-scalar ratio $r=3.3\times 10^{-3}$. 
On the other hand, since the amplitude of scalar perturbation is given as
$A_S=\frac{V_E}{24\pi^2M_{\rm pl}^4\epsilon}$ and the CMB observation constrains
it as $A_S=2.4\times 10^{-9}$ at $k_\ast=0.002~{\rm Mpc}^{-1}$ \cite{cmb},
$\kappa_S$ has to satisfy $\kappa_S=4.7\times 10^{-10}\xi^2$ for $N_e=60$.
Using this constraint, the inflaton mass is found to be determined as 
\begin{equation}
m_{S_R}=4.3\times 10^{10}\left(\frac{\xi}{10^3}\right)
\left(\frac{u}{10^{12}~GeV}\right)~{\rm GeV}.
\end{equation}
The inflaton mass should be fixed in a consistent way with eqs.~(\ref{fa}) and (\ref{vevc}).
We also note that the assumed vacuum with the spontaneous  
$CP$ violation could be consistently realized for suitable parameters 
in this inflation framework.

The reheating after the end of inflation is expected to be caused by 
the inflaton decay to the singlet neutrino pairs $N_{i}N_{i}$ 
through the couplings in eq.~(\ref{lyukawa2}). 
In the case $y^N_3>\tilde y^N_{1,2}$ which is assumed in this study, 
a dominant process is $S_R\rightarrow N_{3}N_{3}$.
Since singlet fermions $N_{i}$ interact with other fields only through
the neutrino Yukawa couplings except for the couplings with $S$ and $S^\ast$, 
instantaneous reheating is expected to occur for the case 
$M_{S_R}>2M_{N_3}$ and $H\simeq\Gamma_S~{^>_\sim}~\Gamma_{N_{3}}$, 
where $\Gamma_S$ and $\Gamma_{N_{3}}$ are the decay width 
of $S_R\rightarrow N_3N_3$ and $N_{3}\rightarrow \bar\ell_i\phi^\dagger$, 
respectively.\footnote{Here, we do not consider a possibility 
for non-thermal leptogenesis which could be expected to occur for the case 
$\Gamma_{N_i}>\Gamma_S$ \cite{nonth}.}
If we take account of these conditions which may be expressed as 
$\sqrt{\kappa_S}>y_3^N$ and $2\sqrt{\kappa_S}y^N_3~{^>_\sim}~ y_3^{\nu 2}$, 
the reheating temperature $T_R$ could be bounded as\footnote{
The restoration of the PQ symmetry could occur 
in the reheating process depending on the parameters. 
However, since the domain wall number is one
in this model, no domain wall problem is induced even if the PQ
symmetry is restored.} 
\begin{equation}
 T_R\simeq 1.6\times 10^8y_3^\nu(y_3^Nu)^{1/2}
< 5\times 10^{11}\left(\frac{\xi}{10^3}\right)^{3/2}
\left(\frac{u}{10^{12}~{\rm GeV}}\right)^{1/2}~{\rm GeV}.
\end{equation}
Although this shows that $T_R>M_{N_1}(\equiv \tilde y_1^Nu)$ could be satisfied
for suitable parameters, $N_1$ is not expected to be thermalized as a
relativistic particle since the Yukawa coupling $y_1^\nu$ of $N_{1}$ 
is supposed to be very small.
Fortunately, it could be expected to reach the thermal 
equilibrium through the scattering process
$N_{3}N_{3}\rightarrow N_{1}N_{1}$ mediated by the scalar $S_R$
before it becomes non-relativistic $(T<M_{N_1})$.
This allows the present model to generate the lepton number 
asymmetry sufficiently through the out-of-equilibrium decay of $N_{1}$
although the Yukawa coupling $y_1^\nu$ of $N_1$ is very small.
We discuss this possibility in the next part.
  
\subsection{Leptogenesis}
The model has two possible decay processes
 $N_1 \rightarrow \ell_i\phi^\dagger$ and 
$\Sigma_\alpha \rightarrow \ell_i\eta^\dagger$
which could contribute to the generation of 
the lepton number asymmetry since 
these processes violate the lepton number.
However, $\Sigma_\alpha$ has the $SU(2)$ gauge interaction
so that its out-of-equilibrium decay is impossible at least 
before the electroweak symmetry breaking. 
On the other hand, as addressed above, 
$N_1$ could reach the equilibrium abundance 
through the scattering mediated by $S_R$ 
even if its neutrino Yukawa coupling $y_1^\nu$ is very small.
In that case, its decay could generate the lepton number asymmetry
through the out-of-equilibrium decay at $T<M_{N_1}$.
As long as the couplings $y_i^N$ or $\tilde y^N_i$ between the inflaton 
and $N_i$ have sufficient magnitude such as 
$y_3^N > \tilde y_2^N~{^>_\sim}~\tilde y_1^N ~{^>_\sim}~10^{-3}$, 
the equilibrium number
density of $N_1$ can be easily realized at $T>M_{N_1}$ 
as shown below.\footnote{We should recall that $y^N_{1,2}=0$ and 
$\tilde y^N_3=0$ are assumed.}
On the other hand, since the mass of $N_i$ is generated through 
the Yukawa coupling $(y_i^NS+\tilde y^N_iS^\ast) \bar N_i^cN_i$, $N_i$ 
cannot be light if we
take account of the values of $y_i^N$ and $\tilde y^N_i$ mentioned above.
In fact, under the constraints (\ref{fa}) and (\ref{vevc}),
the mass of $N_1$ has to be $M_1>10^7$ GeV at least.

In order to check whether this scenario works, we present a typical solution 
of the Boltzmann equations for $Y_{N_1}$ and $Y_{N_3}$ as functions of
$z(\equiv \frac{M_{N_1}}{T})$ in the left panel of Fig.~2.
Here, $Y_{N_i}$ is defined as $Y_{N_i}=\frac{n_{N_i}}{s}$ with the $N_i$ number 
density $n_{N_i}$ and the entropy density $s$. 
In this calculation, as an example, we assume 
$u=2\times 10^{12}~{\rm GeV}$ and $\xi= 500$ 
and then the inflaton mass is fixed as $m_{S_R}=4.3\times 10^{10}$ GeV.
Taking account of the constraints in eq.~(\ref{nmcond}), we fix other relevant 
parameters at the following values,\footnote{These values of $y^\nu_i$ 
require some overall suppression effect compared with the Yukawa couplings 
of the up type quarks in eq.~(\ref{inityuk}). We just assume it in this setting.}
\begin{eqnarray}
&&y_3^N=10^{-2},  \qquad  \tilde y_2^N=10^{-0.2}y_3^N, 
\qquad  \tilde y_1^N= 10^{-0.5}y_3^N
\nonumber \\ 
&&y^\nu_3=2.8\times 10^{-3}, \qquad y_2^\nu= 10^{-5}, \qquad y_1^\nu=10^{-6}.
\label{para1}
\end{eqnarray}
Since $m_{S_R}>2M_{N_3}$ is satisfied, $N_3$ is allowed to be produced 
through the inflaton decay $S_R\rightarrow N_3N_3$. 
The following $N_3$ decay $N_3\rightarrow\ell_i\phi^\dagger$
caused by the coupling $y_3^\nu$ is considered to be a substantial 
process for the thermalization. Thus,
the initial value of $Y_{N_3}$ is fixed as the one produced through this 
inflaton decay assuming the instantaneous reheating. 
The figure shows that $Y_{N_1}$ reaches the equilibrium value 
$Y_{N_1}^{\rm eq}$ around $z\simeq 1$ for the assumed value of $\tilde y_1^N$ 
and leaves its equilibrium value at $z~{^>_\sim}~ 1$ where the out-of-equilibrium decay could generate the lepton number 
asymmetry.\footnote{For a smaller value of $\tilde y_1^N$ or a larger
value of $y^\nu_3$, $Y_{N_1}$ cannot reach an equilibrium value 
$Y_{N_1}^{\rm eq}$ for $z<1$ although $Y_{N_1}$ keeps a constant value 
in the same way as the one in the left panel of Fig.~2.
A larger value of $\tilde y_1^N$ realizes $Y_{N_1}=Y_{N_1}^{\rm eq}$ 
at an earlier stage. }  

\begin{figure}[t]
\begin{center}
\epsfxsize=7.5cm
\leavevmode
\epsfbox{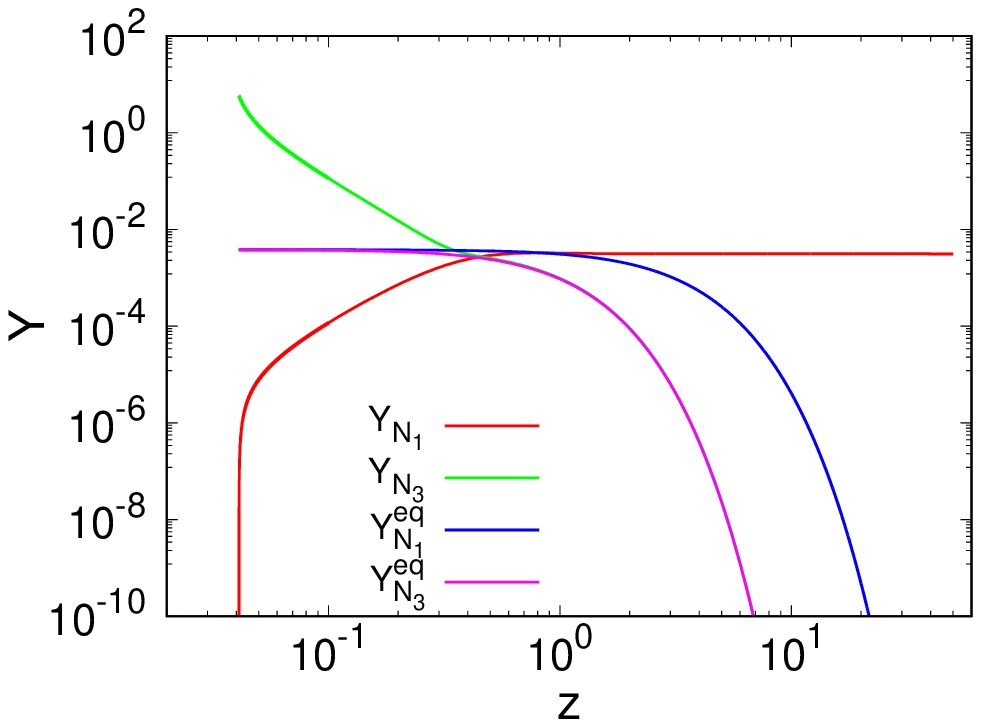}
\hspace*{5mm}
\epsfxsize=7.5cm
\leavevmode
\epsfbox{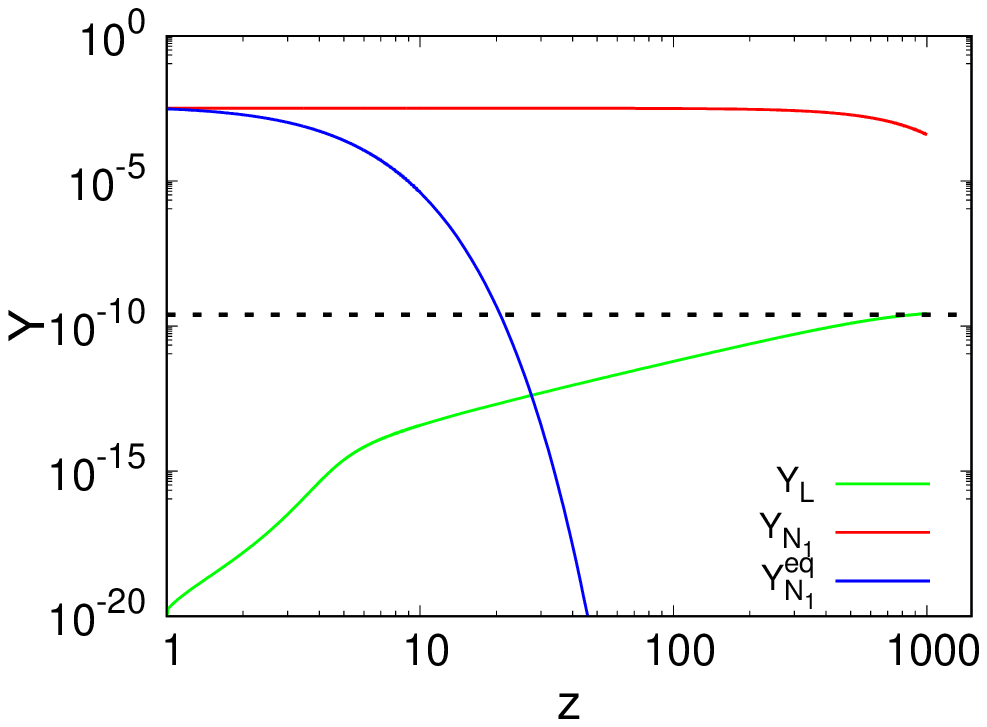}
\end{center}
\vspace*{-3mm}
    {\footnotesize {\bf Fig.~2}~~The left panel: a typical solution of the 
Boltzmann equations for $Y_{N_1}$ and $Y_{N_3}$.
Their equilibrium values $Y_{N_1}^{\rm eq}$ and $Y_{N_3}^{\rm eq}$ are also 
plotted in the same panel. While $Y_{N_3}$ is found to follow $Y_{N_3}^{\rm eq}$ 
at $z~{^>_\sim}~0.4$, $Y_{N_1}$ keeps a constant value until $N_1$ starts decaying.
The right panel: the evolution of the lepton number asymmetry $Y_L$ 
generated through the out-of-equilibrium decay of $N_1$. Horizontal dotted
lines show the value of $Y_L$ required in this model 
to realize the baryon number asymmetry in the Universe \cite{pdg}. }  
\end{figure}  

The generated lepton number asymmetry through the $N_1$ decay 
is converted to the baryon number asymmetry through the sphaleron
process as in the usual leptogenesis \cite{fy,leptg}.
In the present model, the $CP$ asymmetry $\varepsilon$ for  the decay
$N_1 \rightarrow \ell\phi^\dagger$ \cite{cp} is dominantly caused by an 
interference between a tree diagram and a one-loop diagram 
mediated by $N_3$ which are shown in Fig.~3.
Under the assumption given in eq.~(\ref{flavor}), it can be estimated as 
\begin{eqnarray}
  \varepsilon&\equiv&\frac{\Gamma(N_1\rightarrow\ell\phi^\dagger)-
\Gamma(N_1^c\rightarrow\bar\ell\phi)}
{\Gamma(N_1\rightarrow\ell\phi^\dagger)+
\Gamma(N_1^c\rightarrow\bar\ell\phi)} 
=\frac{1}{8\pi}
\frac{{\rm Im}(\sum_i\tilde y^\nu_{1i}e^{-i\frac{\rho}{2}}
y_{3i}^{\nu\ast}e^{-i\frac{\rho}{2}})^2}
 {(\sum_i y_{1i}^\nu y_{1i}^{\nu \ast})}
F\left(\frac{M_{N_3}^2}{M_{N_1}^2}\right)
             \nonumber \\
  &=&\frac{1}{4\pi}
|y^\nu_3|^2F\left(\left(\frac{y_3^{N}}{\tilde y_1^{N}}\right)^2\right)
\sin(-2\rho),   
\end{eqnarray}
where $F(x)$ is defined as
\begin{equation}
F(x)=\sqrt{x}\left(1-(1+x)\ln\frac{1+x}{x}\right).
\end{equation}
In the following analysis, we assume $\sin(-2\rho)=1$ which makes
$\varepsilon$ maximal.

If $N_1$ is in the thermal equilibrium at $z<1$,
the out-of-equilibrium decay of $N_1$ could start at $z\sim 1$ and
the lepton number asymmetry is effectively generated at $z> 1$.  
By introducing an efficiency factor for the washout of the generated lepton 
number asymmetry as $\kappa$,
the lepton number asymmetry $Y_L$ which is defined as
$Y_L\equiv \frac{n_L}{s}$ by using a net lepton number density $n_L$
is roughly estimated as $Y_L =\varepsilon \kappa Y_{N_1}^{\rm eq}|_{z=1}$.
It suggests that $\varepsilon~{^>_\sim}~  8\times 10^{-8}\kappa^{-1}$ is necessary  
to realize a value $Y_L~{^>_\sim}~2.5\times 10^{-10}$ at a sphaleron decoupling 
temperature in order to produce the sufficient baryon number asymmetry in the Universe for $Y_{N_1}^{\rm eq}|_{z=1} \simeq 3.1\times 10^{-3}$.
Since $y_1^\nu$ is supposed to be very small in this model, $N_1$ is 
considered to start its substantial decay at a later stage such as $z\gg 1$ 
where the washout caused by $N_3$ and $\Sigma_\alpha$ 
could be largely Boltzmann suppressed  
as long as $\frac{M_{N_3}}{M_{N_1}}, \frac{M_{\Sigma_\alpha}}{M_{N_1}}> 1$
are satisfied.  Thus, in such a case, the almost all lepton 
number asymmetry generated there could be kept and
the sufficient lepton number asymmetry is expected to be generated through 
the out-of-equilibrium decay of $N_1$.

\begin{figure}[t]
\begin{center}
\epsfxsize=10cm
\leavevmode
\epsfbox{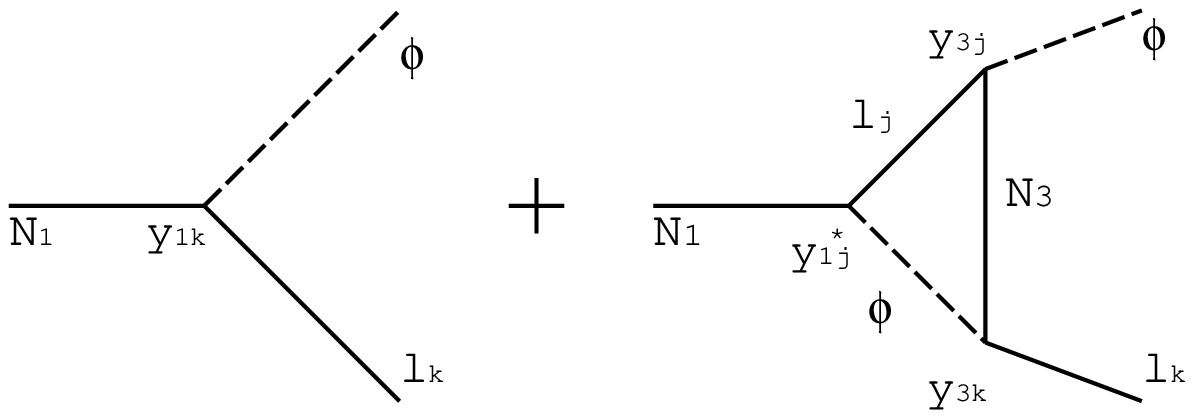}
\end{center}
\vspace*{-5mm}
    {\footnotesize {\bf Fig.~3}~~The $N_1$ decay diagrams which contribute to 
the generation of the lepton number asymmetry. The interference between them
causes the $CP$ asymmetry $\varepsilon$. }  
\end{figure}  

In the right panel of Fig.~2, we present the evolution of the lepton number 
asymmetry $Y_L$ generated through the out-of-equilibrium decay of $N_1$
using the same parameters given in eq.~(\ref{para1}) which can prepare 
an initial value $Y_{N_1}(1)\simeq Y_{N_1}^{\rm eq}(1)$ as shown in the left
panel.
In this analysis of Boltzmann equations, we fully take account of the
washout processes and use the neutrino Yukawa couplings 
$h_{1,2}$ which are fixed by taking account of the condition 
(\ref{nmcond}) with $M_\eta=10^3~{\rm GeV}$, 
$M_{\Sigma_{1,2}}=3M_{N_1}$ and $|\lambda_5|= 10^{-1.5}$.
Since the small neutrino Yukawa coupling $y_1^\nu$ makes the $N_1$ decay 
be delayed to the temperature where the washout processes could be 
frozen out due to the Boltzmann suppression.
This feature can be found in the behavior of $Y_N$ and $Y_L$ in the right panel.   
As its result, almost all the lepton number asymmetry generated through
the out-of-equilibrium $N_1$ decay could be converted to the baryon number 
asymmetry in the Universe as discussed above. 
The model is found to present a successful leptogenesis framework. 
Results of the analysis for several parameter settings are also listed in Table 1. 

Here, we order a few remarks related to these results. 
First, since a smaller $|\lambda_5|$ makes $h_{1,2}$ larger through 
the neutrino mass condition (\ref{nmcond}) for the fixed $M_{\Sigma_{1,2}}$,
the washout processes mediated by $\Sigma_{1,2}$ are considered to suppress 
the generation of the lepton number asymmetry at an early stage 
where it is not frozen out. 
Second, the $N_1$ mass seems to be bounded as $M_1>10^9$ GeV 
in the present model in order to produce the required baryon number 
asymmetry. This bound is similar to the one given in \cite{di}.
Third, for the present parameter settings, $w~{^>_\sim}~10^{10}~{\rm GeV}$ 
seems to be required to avoid the washout of the generated lepton number asymmetry, which is consistent with the requirement from 
the PQ symmetry breaking scale.
Finally, the coexistence of the couplings $y^N_i$ and $\tilde y^N_i$ such as
$y^N_i\not= \tilde y^N_i$ in eq.~(\ref{lyukawa2}) is crucial 
for the leptogenesis.   We should recall that the same feature is 
required in the explanation of the
CKM phase through the mass matrix (\ref{dmass}).

\begin{figure}[t]
\footnotesize
\begin{center}
\begin{tabular}{cc|cccc}\hline
$M_{\Sigma_{1,2}}$ & 
$|\lambda_5|$ &  $h_1$ & $h_2$ & $|\varepsilon|$ & $Y_B$  
\\ \hline 
 $3M_{N_1}$ &  $10^{-1}$  & $3.7\times 10^{-2}$ & $1.3\times 10^{-2}$
&$9.6\times 10^{-8}$     &  $9.5\times 10^{-11}$  \\
 $3M_{N_1}$ &  $10^{-1.5}$ & $6.6\times 10^{-2}$  & $2.4\times 10^{-2}$ 
 &$9.6\times 10^{-8}$     &  $9.4\times 10^{-11}$  \\
$3M_{N_1}$ &  $10^{-2}$  & $1.2\times 10^{-1}$ &
$4.2\times 10^{-2}$ &  $9.6\times 10^{-8}$    &  $7.2\times 10^{-11}$  \\ \hline
$5M_{N_1}$ &  $10^{-1.5}$    &$8.4\times 10^{-2}$  & $3.0\times 10^{-2}$ &  $9.6\times 10^{-8}$ &
$9.4\times 10^{-11}$  \\
$10M_{N_1}$ &   $10^{-1.5}$ &   $1.2\times 10^{-1}$ &
$4.2\times 10^{-2}$ &  $9.6\times 10^{-8}$    
&$9.4\times 10^{-11}$   \\ 
\hline
\end{tabular}
\end{center}
\normalsize
\footnotesize{{\bf Table 1}~ 
The $CP$ asymmetry $\varepsilon$ and the generated baryon 
number asymmetry $Y_B$ for the parameters in eq.~(\ref{para1})
with $u=2\times 10^{12}$ GeV and $\xi=500$, which realize 
the spectrum $\frac{m_{S_R}}{2}>M_{N_3}>M_{N_2}>M_{N_1}$.
The Yukawa couplings $h_{1,2}$ of $\Sigma_{1,2}$ are determined 
through the neutrino oscillation conditions (\ref{nmcond}) by assuming 
the values of $|\lambda_5|$ and $M_{\Sigma_{1,2}}$.}
\end{figure}
 
\subsection{Dark matter}
The model has three dark matter (DM) candidates, that is, the axion,
the neutral component of $\Sigma_\alpha$ and
the lightest neutral component of $\eta$.
The axion could explain the required DM abundance as long as
$w\simeq 10^{12}$~GeV is satisfied \cite{fa}.
The latter two have odd parity of the remnant $Z_2$ of the global 
$U(1)$ symmetry, which makes them stable and then DM candidates.
However, $\Sigma^0_\alpha$ is supposed to
have a large mass so that it cannot be DM in the present
model.\footnote{The DM study in the cases where $\Sigma$ has 
a mass of $O(1)$ TeV can be found in \cite{scot3}.}
On the other hand, $\eta$ is assumed to have a mass of $O(1)$~TeV as
discussed in the neutrino mass generation.
In that case, the lightest neutral component of $\eta$ can be DM.
Moreover, even if the VEV $w$ is not large enough to guarantee
the sufficient axion density for the explanation of the DM energy density,
the thermal relics of $\eta^0$ could explain it as long as the quartic
couplings $\lambda_{3,4}$ in eq.~(\ref{spot}) take suitable values \cite{etadm,ks}.
As a result, the breaking scale $w$ of the PQ symmetry could be free
from the explanation of the DM energy density in this model.
  
\subsection{Quark and lepton mass hierarchy}
Yukawa coupling constants for quarks and leptons are
  related each other by eq.~(\ref{inityuk}) at an $SU(4)$ breaking
  scale $\Lambda$.
On the other hand, their weak scale values which determine mass 
eigenvalues of the quarks and the leptons are fixed through the 
renormalization group equations taking them as the initial values.
It can bring about a difference of a factor three 
due to the color effect between quarks and leptons.
The mass difference between the down type quarks and the charged 
leptons seems to be partially explained by this effect but 
it is not satisfactory. 
Even if corrections caused by the mixing with heavy fermions 
in these sectors are taken into account,
this situation is not improved and then some new ingredients 
are needed to be introduced for it.

On the other hand, in the up type quarks and the neutrinos,
several additional parameters related to the neutrino mass generation
could give a different feature in these sectors. Especially, since
neutrino masses are determined by the type III seesaw contribution,
the relation among the Yukawa couplings of quarks and leptons 
at the high energy scale is not directly affect their mass matrices.
These features could make the large difference found in the CKM and 
PMNS matrices be consistently realized in the present unification scheme.
Since details depend on the model parameters and this issue is 
beyond the scope of present study, we will not
discuss it further here and leave it to future study.
Finally, it may be useful to note the fact that the present unification scheme
could make the leptogenesis work well.
A requirement that the third generation Yukawa coupling of the up quark 
sector should be much larger than others brings about the relation 
$y_{1,2}^\nu\ll y_3^\nu$ in the neutrino sector, 
which plays a crucial role in the present leptogenesis scenario
as shown in the above study 

\section{Summary}  
We proposed a model which gives the origin of the $CP$ violation
at an intermediate scale. In this model, the $CP$ symmetry is supposed to be
spontaneously broken but it does not cause the strong $CP$ problem
and $\bar\theta=0$ is kept even if the radiative corrections are 
taken into account. We showed that such a model could be
realized in a Pati-Salam type unification model, in which $CP$ phases in both the
CKM and PMNS matrices are derived from the same source.
Neutrino masses are generated in a hybrid way by the tree level type 
I seesaw and the one-loop type III seesaw. 
The required baryon number asymmetry can be produced through
the leptogenesis. The out-of-equilibrium decay of $N_1$ occurs 
at a later stage where the washout effects are almost frozen out.
As a result,
the generated lepton number asymmetry could be effectively converted
to the baryon number asymmetry.
This feature comes from the present unification based on a fact that 
the top Yukawa coupling is much larger than others.
The model has two DM candidates and
the dominant DM is fixed depending on the intermediate symmetry
breaking scale. Since the axion needs not to be DM,  
the PQ symmetry breaking scale can be free from the condition for the
DM energy density realization.  
We also note a possibility such that the model might be derived as 
the low energy effective model of the $E_8\times E_8^\prime$ superstring.
It will be discussed elsewhere.

\section*{Appendix}
In this Appendix, we present a simple example which could bring 
about a phase in the CKM matrix.
We assume the relevant couplings $y^d$, $y^D$ and
$\tilde y^D$ to be written as\footnote{A similar Yukawa coupling matrix
for the down type quarks has been considered in a different context
\cite{flavor}. There is no background to explain
its hierarchical structure in the present model.}
\begin{equation}
y^d=c\left(\begin{array}{ccc}
\epsilon^4 &\epsilon^3 &x\epsilon^3\\
\epsilon^3 &\epsilon^2 & y\epsilon^2 \\
\epsilon^2 & 1 & -1 \\
\end{array}\right), \quad
y^D=(a_1, a_2, a_3), \quad \tilde y^D=(b_1, b_2, b_3),
\end{equation}
by using real constants $a_i$, $b_i$, $c$ and $x,~y$.
As long as $\epsilon$ satisfies $\epsilon \ll 1$, 
the down type quark mass matrix $m^d(\equiv y^d\langle\tilde\phi\rangle)$ 
has hierarchical mass eigenvalues. 
Here, we introduce $X_{ij}$ and $Y_{ij}$ whose definition is given as
\begin{eqnarray}
X_{ij}&=&1+p_ip_j  \nonumber \\
&+&\frac{(a_2+b_2)^2+(a_3+b_3)^2p_ip_j
+\{a_2b_3+b_2b_3+ (a_2b_3+a_3b_2)\cos 2\rho\}(p_i+p_j)}
{a_2^2+a_3^2+b_2^2+b_3^2+2(a_2b_2+a_3b_3)\cos 2\rho}, \nonumber \\
Y_{ij}&=&\frac{(a_2b_3-a_3b_2)(p_i-p_j)\sin 2\rho}
{a_2^2+a_3^2+b_2^2+b_3^2+2(a_2b_2+a_3b_3)\cos 2\rho}
\end{eqnarray}
where $p_i$ is fixed as $p_1=x,~p_2=y$ and $p_3=-1$.
If we define $R_{ij}$ and $\theta_{ij}$ by using these quantities as
\begin{equation}
R_{ij}=\sqrt{X_{ij}^2+Y_{ij}^2}, \quad \tan\theta_{ij}=\frac{Y_{ij}}{X_{ij}},
\end{equation}
the component of eq.~(\ref{ckm}) is found to be expressed as
\begin{equation}
(A^{-1}m^2A)_{ij}=c^2\langle\tilde\phi\rangle^2 \epsilon_{ij}R_{ij}e^{i\theta_{ij}},
\label{diag}
\end{equation}
where $\mu_D^2\ll {\cal F}^d{\cal F}^{d\dagger}$ is assumed.
$\epsilon_{ij}$ is defined as
\begin{equation}
\epsilon_{11}=\epsilon^6, \quad \epsilon_{22}=\epsilon^4, \quad
\epsilon_{33}=1, \quad \epsilon_{12}=\epsilon_{21}=\epsilon^5, \quad
\epsilon_{13}=\epsilon_{31}=\epsilon^3, \quad
\epsilon_{23}=\epsilon_{32}=\epsilon^2.
\end{equation}

By solving eq.~(\ref{diag}), we find that $A$ is approximately written as
\begin{equation}
A\simeq\left(
\begin{array}{ccc}
1 & -\lambda & \lambda^3\left(\frac{X_{23}}{|\alpha|^2X_{33}}e^{i\theta}
-\frac{X_{13}}{|\alpha|^3X_{33}}\right) \\
\lambda & 1 & -\lambda^2 \frac{X_{23}}{|\alpha|^2X_{33}}e^{i\theta}\\
 \lambda^3\frac{X_{13}}{|\alpha|^3X_{33}}&
 \lambda^2\frac{X_{23}}{|\alpha|^2X_{33}}e^{-i\theta} & 1 \\
\end{array}
\right),
\end{equation}
where the constants $\lambda$, $\alpha$ and $\theta$ are defined as
\begin{equation}
\alpha=\frac{X_{12}X_{33}-X_{13}X_{23}
e^{-i(\theta_{23}+\theta_{12}-\theta_{13})}}
{X_{22}X_{33}-X_{23}^2}, \quad
\lambda=|\alpha|\epsilon,\quad
\theta=\arg(\alpha) +\theta_{23}+\theta_{12}-\theta_{13}. 
\end{equation}
This expression shows that $A$ could have a non-trivial phase which 
gives the origin of the CKM phase as long as $a_2b_3-a_3b_2\not=0$ and 
$x\not=y$ are satisfied. If the diagonalization matrix $O^L$ for the 
mass matrix of the up type quarks takes an almost diagonal form,
an interesting matrix could be obtained as the CKM matrix such as 
$V_{CKM}\simeq A$. 
In this case, the mass eigenvalues for the down type quarks are obtained as
\begin{eqnarray}
&&X_{33}^{1/2}c\langle\tilde\phi\rangle, \qquad
\left(X_{22}-\frac{X_{23}^2}{X_{33}}\right)^{1/2}\epsilon^2c\langle\tilde\phi\rangle, 
\nonumber \\
&&\left\{X_{11}-\frac{X_{13}^2}{X_{33}}+|\alpha|^2\left(X_{22}-\frac{X_{23}^2}{X_{33}}-2\right)\right\}^{1/2}\epsilon^3c\langle\tilde\phi\rangle.
\end{eqnarray} 

A diagonalization matrix $\tilde A$ for the charged lepton mass matrix takes
the same form as $A$ as a result of the Pati-Salam $SU(4)$ symmetry 
in the model. However, since the Yukawa couplings which induce the neutrino 
mass matrix could be irrelevant to the ones in the up type quarks 
as discussed in the text, the large mixing in the PMNS matrix could be 
obtained if large flavor mixings are realized 
in the neutrino mass matrix. If we use the assumption in eq.~(\ref{flavor}),
the PMNS matrix in this example is found to be written as
\begin{equation}
V_{PMNS}=\left(\begin{array}{ccc}
\frac{1}{\sqrt 6}(2-\lambda) & \frac{1}{\sqrt 3}(1+\lambda) &
\frac{1}{\sqrt 2}\lambda \\
\frac{1}{\sqrt 6}(-1-2\lambda+\beta\lambda^2) & 
\frac{1}{\sqrt 3}(1-\lambda-\beta\lambda^2) &
\frac{1}{\sqrt 2}(1+\beta\lambda^2) \\
\frac{1}{\sqrt 6}(1+\beta^\ast\lambda^2) & 
-\frac{1}{\sqrt 3}(1+\beta^\ast\lambda^2) &
\frac{1}{\sqrt 2}(1-\beta^\ast\lambda^2) \\
\end{array}
\right) +O(\lambda^3),
\end{equation}
 where $\beta=\frac{X_{23}}{|\alpha|^2X_{33}}e^{i\theta}$ and the Majorana phases
are not taken into account.

\section*{Acknowledgements}
This work is partially supported by MEXT Grant-in-Aid 
for Scientific Research on Innovative Areas (Grant No. 26104009)
and a Grant-in-Aid for Scientific Research (C) from Japan Society
for Promotion of Science (Grant No. 18K03644).

\newpage
\bibliographystyle{unsrt}

\end{document}